\documentclass[journal=nalefd,manuscript=letter]{achemso}

\setkeys{acs}{journal=nalefd}

\usepackage{amsmath}
\usepackage{graphicx}
\usepackage{booktabs}
\usepackage{siunitx}
\usepackage{fix-cm}  
\usepackage{etoolbox}
\usepackage{ulem}
\usepackage{soul}
\usepackage{pdfpages}
\PassOptionsToPackage{normalem}{ulem}
\AtEndDocument{\clearpage\csname @tocentry\endcsname}
\author{Vinod~K.~Paidi}
\email{vinod-kumar.paidi@insa-lyon.fr}
\affiliation{CNRS, INSA Lyon, UCBL, MATEIS, UMR 5510, 69621 Villeurbanne, France}

\author{Yves~Joly}
\affiliation{Université Grenoble Alpes, CNRS, Institut Néel, F38042 Grenoble, France}

\author{Alex~Taekyung~Lee}
\affiliation{Department of Chemical Engineering, University of Illinois at Chicago, Chicago, IL 60608, United States}
\alsoaffiliation{Materials Science Division, Argonne National laboratory, Lemont, IL 60439, United States}

\author{Guillaume~Radtke}
\affiliation{Sorbonne Université, CNRS UMR 7590, MNHN, IMPMC, 4 place Jussieu, 75252 Paris, France}

\author{Vlado~K.~Lazarov}
\affiliation{SuperSTEM Laboratory, SciTech Daresbury Campus, Daresbury WA4 4AD, United Kingdom}
\alsoaffiliation{York-JEOL Nanocentre, University of York, United Kingdom}

\author{Demie~Kepaptsoglou}
\affiliation{SuperSTEM Laboratory, SciTech Daresbury Campus, Daresbury WA4 4AD, United Kingdom}
\alsoaffiliation{School of Physics, Engineering and Technology, University of York, United Kingdom}

\author{Quentin~M.~Ramasse}
\affiliation{SuperSTEM Laboratory, SciTech Daresbury Campus, Daresbury WA4 4AD, United Kingdom}
\alsoaffiliation{School of Physics and Astronomy, University of Leeds, United Kingdom}
\alsoaffiliation{School of Chemical and Process Engineering, University of Leeds, United Kingdom}

\author{Matthieu~Bugnet}
\email{matthieu.bugnet@insa-lyon.fr}
\affiliation{CNRS, INSA Lyon, UCBL, MATEIS, UMR 5510, 69621 Villeurbanne, France}
\alsoaffiliation{SuperSTEM Laboratory, SciTech Daresbury Campus, Daresbury WA4 4AD, United Kingdom}
\alsoaffiliation{School of Chemical and Process Engineering, University of Leeds, United Kingdom}

\title{Site-Specific Orbital Symmetry Breaking in Epitaxial Graphene on Silicon Carbide}


\begin{document}
\clearpage
\begin{abstract}
The epitaxial graphene grown on SiC interface has been the subject of tremendous research efforts to control its electronic transport properties in the quest for graphene electronic applications. The electronic structure of the epitaxial graphene buffer layer on SiC(0001) is investigated at the atomic scale using monochromated electron energy-loss spectroscopy (EELS) in an aberration-corrected scanning transmission electron microscope (STEM). The C $K$-edge reveals spectral fine structures distinctly different from free-standing graphene layer, providing direct experimental evidence of a modified bonding environment. \textit{Ab initio} calculations providing a site-resolved probe of orbital character to elucidate the origin of these spectral features: covalent bonding between one-quarter of the buffer-layer carbon atoms and the substrate silicon atoms induces partial $sp^3$ hybridization, quenching the out-of-plane $\pi$ states and disrupting $\pi$-conjugation across the interfacial layer. Substrate-imposed symmetry breaking further lifts the in-plane orbital degeneracy. This orbital-engineering framework, rooted in $\pi$-state quenching and site-specific symmetry breaking, establishes a quantitative basis for tailoring graphene–substrate interfaces toward next-generation low-energy transistor applications.
\end{abstract}

\section{Introduction}

As silicon-based device technology approaches its fundamental physical scaling limits, epitaxial graphene on SiC has emerged as a high-performance candidate for next-generation nanoelectronics\cite{Zhao2024}. Although pristine graphene is a zero-bandgap semimetal, the epitaxial layer in direct contact with the (0001)-SiC substrate — commonly designated the buffer layer (BL)\cite{Varchon2007} — exhibits semiconducting character with a reported bandgap of approximately 0.6~eV and hole mobilities surpassing those of germanium, properties that render it particularly suited to logic transistor applications\cite{Zhao2024}. The synthesis of high-quality graphene by thermal decomposition of hexagonal SiC has further consolidated this system as a viable nanoelectronic platform,\cite{Emtsev2009,Berger2004,Berger2006} motivated by the prospects of nanoribbon-defined bandgap engineering\cite{Sprinkle2010,Hicks2013} and high electron mobility.\cite{Baringhaus2014} The BL is further characterized by a structural reconstruction,\cite{Forbeaux1998} intrinsic $n$-type doping from Si dangling bonds,\cite{Sonde2009} and partial covalent bonding to the underlying Si-terminated SiC surface.\cite{Varchon2007,Riedl2007} Decoupling of the BL via intercalation of H, O, F, or Li,\cite{Riedl2009,Speck2011,Oliveira2013,Wong2011,Virojanadara2010} or rapid thermal quenching,\cite{Bao2016} yields quasi-free-standing graphene with tunable carrier type and concentration, enabling the engineering of novel two-dimensional heterostructures.\cite{Briggs2019} Nevertheless, the relationship between the electronic structure of the BL and its atomic configuration on the (0001) SiC surface remains incompletely understood, representing a fundamental obstacle to the industrial application of this material.

Atomic-scale core-level electron energy-loss spectroscopy (EELS) in the aberration-corrected scanning transmission electron microscope (STEM) is uniquely suited to addressing this problem \cite{pennycook2011scanning}. It offers electronic structure sensitivity at the spatial resolution of a single atomic plane or column,\cite{Muller2008} which provides relevant insight on, e.g., the oxidation state of metals and rare earths \cite{tan20112d, Teurtrie2019}, the charge-carrier distribution in superconductors \cite{gauquelin2014atomic, di2015high, Bugnet2016}, the atomic-site distortion at surfaces and interfaces \cite{torres2011spectroscopic, zhu2012bonding}, etc., and enables mapping orbital directionality \cite{loffler2017real, Bugnet2022, bugnet2024current, iwashimizu2025anisotropic}. 
Previous investigations by Nicotra \textit{et al.}\cite{Nicotra2013} and Palacio \textit{et al.}\cite{Palacio2015} established atomic-scale differentiation of C-$K$ EELS between the BL and adjacent graphene layers, reporting enhanced $\sigma$-bonding character and additional spectral weight between the $\pi^{\ast}$ and $\sigma^{\ast}$ peaks in the BL. However, the instrumental energy resolution employed in these studies ($\sim$0.3 eV) likely precluded the resolution of finer spectral features intrinsic to the BL. Recent interpretation of the C-$K$ electron energy-loss near-edge structure (ELNES) at the graphene–SiC interface have relied on linear combinations of  bulk SiC and free-standing graphene\cite{Gruschwitz2019}, an approach that overlooks the distinct electronic identity of the BL. Lacking theoretical simulations, this methodology cannot resolve how the interface geometry collapses the Dirac cone to form a bandgap. 

In this Letter, epitaxially grown graphene multilayers on (0001) 6H-SiC are investigated using a monochromated electron beam at a $\sim$100 meV energy resolution. Core-loss EELS measurements of the carbon K-edge, acquired from both the BL and the quasi-free-standing graphene, are combined with full-potential finite-difference method near-edge structure (FDMNES) simulations to elucidate the role of in-plane orbital symmetry in the observed bandgap. The C $K$-edge spectra reveal a pronounced spectroscopic contrast between the BL and the overlying quasi-free-standing graphene. The distinct orbital character resolved in the projected density of states demonstrates the broken symmetry at carbon sites non-bonded to SiC substrate, providing a rigorous electronic-structure basis for the deviation of the BL from pristine graphene and, consequently, for the emergence of the substrate-mediated bandgap.

\section{Results and Discussion}
\begin{figure}
  \centering
  \includegraphics[width=\columnwidth]{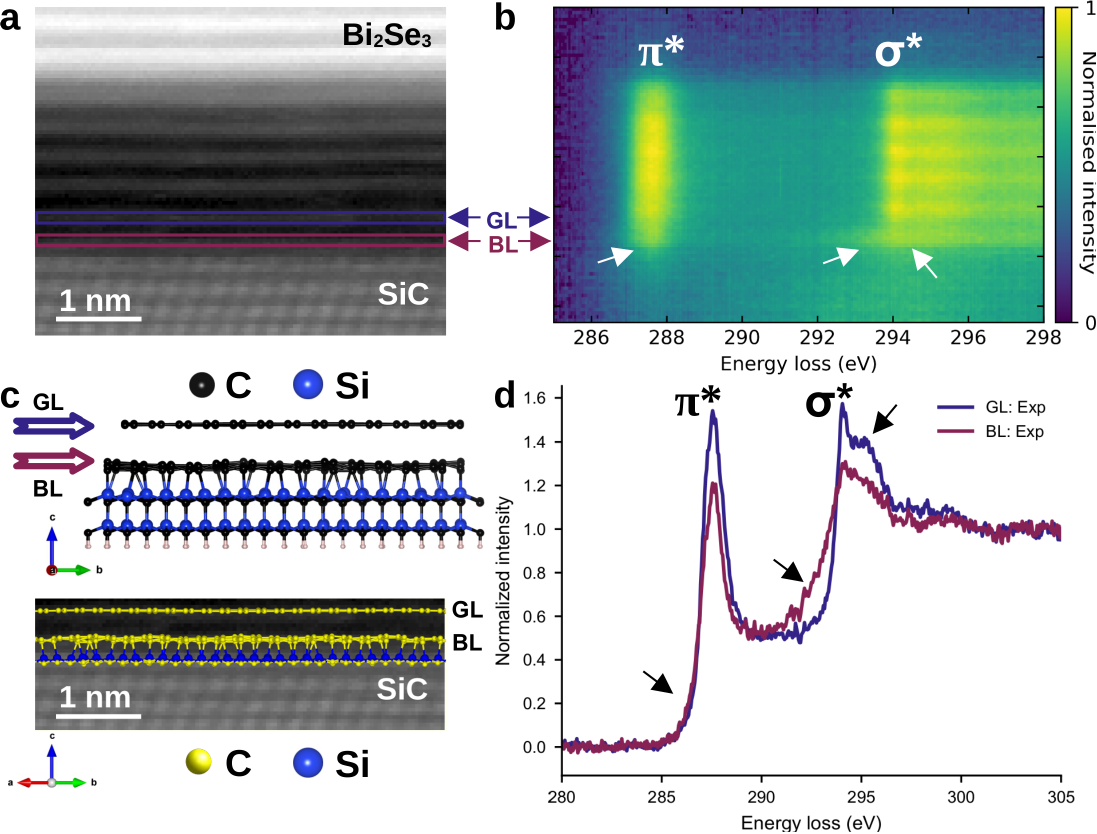}
  \caption{\textbf{Graphene/SiC specimen and layer-by-layer core-loss electron energy loss spectroscopy.} (a) High-resolution HAADF-STEM cross-sectional view of the graphene/SiC specimen. (b) HAADF-STEM image of the graphene/SiC interface acquired as part of the spectrum image. C K-edge ELNES projected along the layer-stacking direction, in which the ELNES of each pixel is summed over the corresponding horizontal line of the datacube. (c) Representative unit cell of the $(6\sqrt{3} \times 6\sqrt{3})\mathrm{R}30^{\circ}$ 
surface reconstruction illustrating the graphene layer (GL) and buffer layer (BL) configurations. Superposition of the relaxed structural model onto the cross-sectional HAADF-STEM image to illustrate the SiC–buffer layer interfacial distance. (d) C K-edge energy-loss near-edge structures of the individual layers: the buffer layer and the first graphene layer.}
  \label{fig:figure1}
\end{figure}

 We investigate a specimen comprising five free-standing graphene layers and an epitaxial buffer layer (BL) grown on a single-crystalline (0001) 6H-SiC substrate, capped with the topological insulator Bi$_{2}$Se$_{3}$ (see Supporting Information), as illustrated by the STEM-high-angle annular dark-field (HAADF) image in Figure~\ref{fig:figure1}(a). The projected C $K$-edge and corresponding HAADF image acquired at the graphene/SiC interface are displayed in Figure~\ref{fig:figure1}(a,b). In the free-standing graphene layers, the $\pi^{\ast}$ feature is clearly resolved at ${\sim}\SI{287.5}{\eV}$, while the $\sigma^{\ast}$ band spans the range \SIrange{293.5}{296}{\eV}. The sharp excitonic resonance at ${\sim}\SI{294}{\eV}$, characteristic of $sp^{2}$-hybridised carbon,\cite{Sato_2016} is well localised on each individual graphene layer\cite{Bugnet2022}. Since the layers above the BL exhibit nearly identical near-edge fine structures, the following discussion focuses on the electronic structure of the first graphene layer (see Figure~\ref{fig:figure1}(c)), hereafter referred to as the graphene layer (GL), for clarity and conciseness. A comparison between the C-$K$ ELNES of the GL and the BL is shown in Figure~\ref{fig:figure1}(d) and Supporting Information, Figure S1; the BL spectrum differs strongly from that of the GL at energies below \SI{297}{\eV}, while the intensity of the continuum of states beyond this energy is comparable. The $\pi^{\ast}$ and $\sigma^{\ast}$ peaks are less intense in the BL than in the GL. Furthermore, Figure~\ref{fig:figure1}(d) reveals additional spectral features, indicated by arrows, most notably an intensity contribution in the energy range \SIrange{288.5}{293.5}{\eV}, situated between the $\pi^{\ast}$ and $\sigma^{\ast}$ resonances, which is likewise apparent in the 2D spectral map presented in Figure~\ref{fig:figure1}(b). The $\sigma^{\ast}$ peak intensity decreases, as does the excitonic peak at ${\sim}\SI{294}{\eV}$, which is attributed to a core-excitonic resonance\cite{Palacio2015, Kepaptsoglou_2015}. The reduced intensity of the BL feature relative to the GL reflects changes in core-excited-state screening arising from covalent bonding at the BL–SiC interface, together with local structural and chemical disorder. Quantifying these experimentally observed excitonic effects (unlike disorder) requires going beyond the single-particle framework employed in our current FDMNES simulations. 
 
Further inspection of the experimental spectra reveals a subtle intensity increase just above the Fermi level, at the onset of the $\pi^{\ast}$ peak. Several of these spectral features are consistent with those present in previously reported spectra,\cite{Nicotra2013,Palacio2015,Gruschwitz2019} though a comprehensive theoretical interpretation has thus far been lacking. To elucidate the origin of each spectral signature specific to the BL, FDMNES simulations of the C\,$1s \rightarrow 2p$ excitation, governed by the dipole-selection rule ($\Delta l = \pm 1$), were performed  using the finite difference method (FDM), which employs a free-shape potential without the spherical approximation (see Supporting Information, Methods), thereby maximizing accuracy in the near-edge region\cite{Yves_2001, Bunau_2009}. The simulations adopted the large realistic (6$\sqrt{3}$~$\times$~6$\sqrt{3}$)$R30^\circ$ surface reconstruction, a structural model that has recently demonstrated improved accuracy in describing the electronic\cite{Cavallucci2018}, vibrational, and thermal properties of the BL.\cite{Radtke2025} As a representative case, side views of the GL and BL atomic structures, together with the superposition of the relaxed structural model onto the cross-sectional HAADF-STEM image, are presented in Figure~\ref{fig:figure1}(c), with full details on the relaxed structural models provided in the Supporting Information, Methods. 

\begin{figure}
  \centering
  \includegraphics[width=\columnwidth]{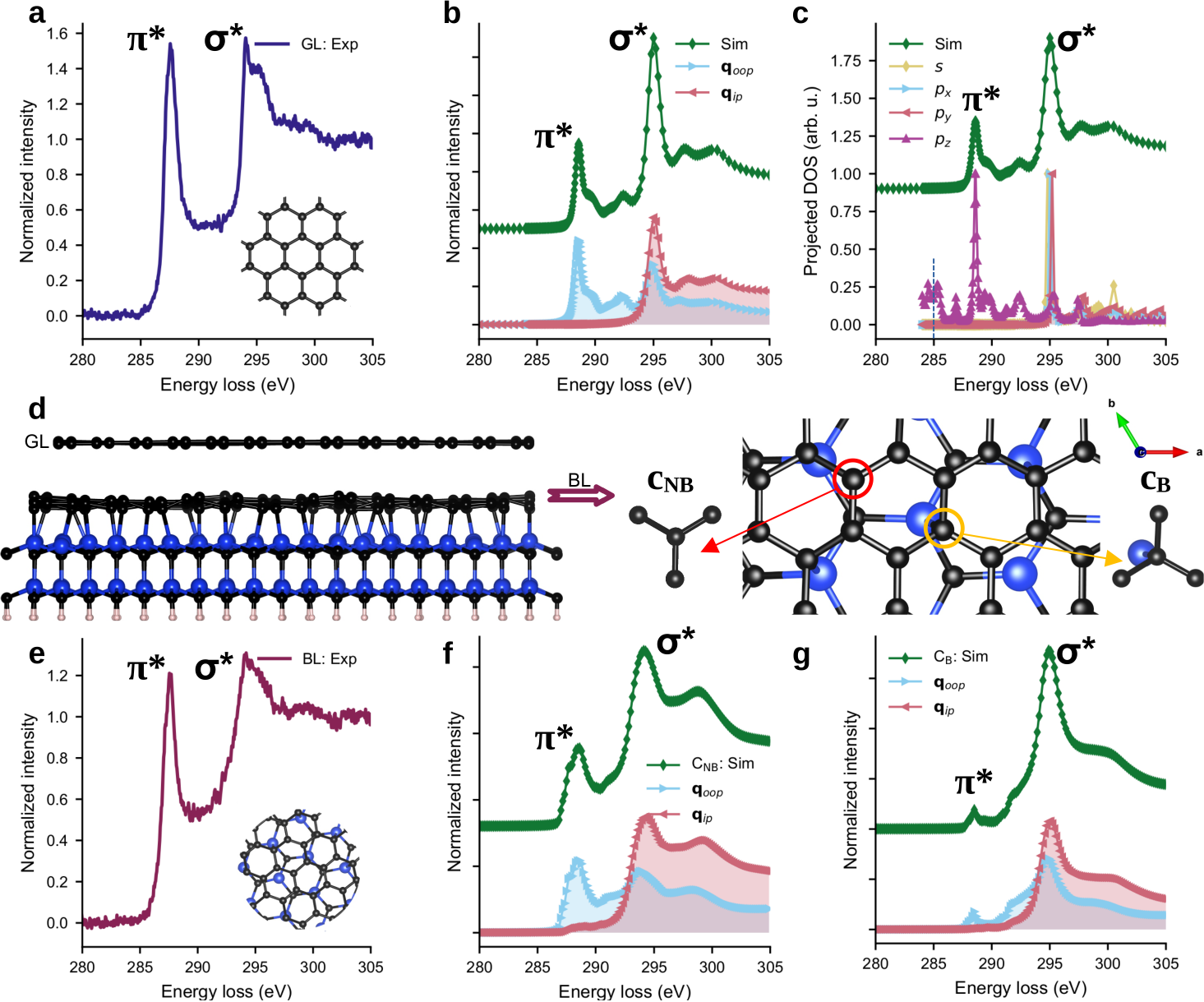}
  \caption{\textbf{C~K-edge ELNES and Simulations of GL and BL on SiC.} (a) Experimental C K-edge ELNES spectrum of the graphene layer (GL). (b) Calculated in-plane ($\mathbf{q}_{ip}$) and out-of-plane ($\mathbf{q}_{oop}$) spectral components (shaded), together with their orientationally weighted sum. (c) Calculated partial density of states (DOS) of GL, with the degenerate $p_x$ and $p_y$ contributions shaded for clarity and shown alongside the calculated spectrum. Dashed line denotes the Fermi level ($E_{\text{F}}$). An energy threshold ($E_{\mathrm{cut}}$) was applied to select transitions into unoccupied states above $E_{F}$ (see Computational Details). (d) Schematic of the GL/SiC interface distinguishing non-bonded ($\mathbf{C}_\text{NB}$) and bonded ($\mathbf{C}_\text{B}$) carbon atoms (black spheres, carbon; blue spheres, silicon). (e) Experimental C K-edge ELNES spectrum of the buffer layer graphene (BL). Calculated C K-edge ELNES and the decomposed $\mathbf{q}_{ip}$ and $\mathbf{q}_{oop}$ components for the (f) $\mathbf{C}_\text{NB}$ and (g) $\mathbf{C}_\text{B}$ sites of BL. }
  \label{fig:figure2}
\end{figure}

Prior to modeling the structurally complex BL, simulations were first performed on GL as a benchmark to validate the observed spectral features. In the experimental geometry employed here (see Supporting Information, Methods, Figure~S2, and Ref.~\cite{Bugnet2022}), the scattering vector $\textbf{q}$ (in units where $\hbar=1$) is a superposition of in-C-plane ($\textbf{q}_{ip}$) and out-of-C-plane ($\textbf{q}_{oop}$) components, with $\textbf{q}_{oop}$ normal to the graphene basal plane, such that the measured C~$K$-edge spectrum (Figures~\ref{fig:figure2}(a) and~\ref{fig:figure2}(e)) reflects a weighted integration over both orientational contributions. The simulations explicitly decompose this anisotropic response: the $\textbf{q}_{ip}$ channel, involving transitions into unoccupied $\sigma^*$ states, and the $\textbf{q}_{oop}$ channel, dominated by transitions into the $\pi^*$ manifold with additional $\sigma^*$ contributions, are computed independently and shown in Figure~\ref{fig:figure2}(b). A linear combination of the two geometric components ($0.44\,\textbf{q}_{oop} + 0.56\,\textbf{q}_{ip}$), following the formalism of H\'ebert et al.~\cite{Hebert2006} (see Supporting Information, Methods), reproduces the experimental spectrum. This assignment is further corroborated by the partial density of states ($p$DOS) projected onto the carbon $p$ orbitals shown alongside the spectra. The sharp onset resonance is unambiguously attributed to pure $p_{z}$ states, corresponding to the $\pi^{*}$ resonance characteristic of graphene, while the broader $\sigma^{*}$ weight at higher energy losses primarily originates from $s$, $p_{x}$, and $p_{y}$ derived states. Notably, as shown in Figure~\ref{fig:figure2}(c), the degeneracy of the $p_{x}$ and $p_{y}$ orbitals is largely preserved. 
These simulations are fully consistent with the canonical $sp^{2}$ hybridization of the graphene lattice, in agreement with previous results\cite{Jorissen_2010, Kepaptsoglou_2015, Junqing_2015, Sato_2016}, providing a clear spectroscopic fingerprint for experimentally observed GL spectra.

Having established a spectroscopic baseline for the GL, the simulation was extended to the more structurally complex BL. In the BL, approximately one quarter of the carbon atoms form covalent bonds with the underlying Si atoms of the SiC substrate (Figure~\ref{fig:figure2}(d)). This partial covalent bonding breaks the lateral homogeneity of the BL, giving rise to two electronically inequivalent carbon sites: those covalently bonded to Si (hereafter $\mathbf{C}_{\mathrm{B}}$) and those remaining unbonded (hereafter $\mathbf{C}_{\mathrm{NB}}$), where $\mathbf{C}_{\mathrm{NB}}$ sites retain a local chemical environment analogous to that of freestanding graphene. To capture this structural heterogeneity, ELNES calculations were performed independently for the carbon sites within the BL. Although the reconstruction possesses no exact symmetry, considering only the immediate local environment of each atom reveals an approximate threefold symmetry, which was exploited to reduce the 338 inequivalent carbon sites to 70 representative sites (See Supporting Information, Methods, and Figure S3) requiring explicit calculation. Each of the 70 calculated site spectra was itself constructed as a linear combination of the two geometric components, $\textbf{q}_{ip}$ and $\textbf{q}_{oop}$, following the same procedure applied to the GL reference. The resulting site-resolved spectra were then weighted and summed to obtain the total simulated response, explicitly accounting for the core-level shifts and the anisotropic momentum-transfer contributions from the $\textbf{q}_{ip}$ and $\textbf{q}_{oop}$ channels at each of the corresponding $\mathbf{C}_{\text{NB}}$ and $\mathbf{C}_{\text{B}}$ sites (Supporting Information, Figure~S4). The site-resolved simulation of the $\mathbf{C}_{\mathrm{NB}}$ sites (Figure~\ref{fig:figure2}(f)) reveals that the $\pi^{*}$ resonance is predominantly governed by the out-of-plane $\textbf{q}_{oop}$ channel, consistent with the purely $p_z$-derived character of the $\pi$ states in an $sp^{2}$-hybridized carbon network. The $\sigma^{*}$ resonance, by contrast, shows contributions from both the $\textbf{q}_{oop}$ and $\textbf{q}_{ip}$ channels, reflecting the mixed in-plane and out-of-plane character of the unoccupied $\sigma$-derived states. This channel decomposition is in good agreement with the anisotropic response established for the GL and confirms that the  $\mathbf{C}_{\mathrm{NB}}$ site retains, to a first approximation, the spectroscopic character of a GL-like carbon environment.

Nevertheless, a comparison between the $\mathbf{C}_{\mathrm{NB}}$ spectrum and the GL reference (see Figures~\ref{fig:figure2}(b) and~\ref{fig:figure2}(f)) reveals a systematic enhancement of the spectral intensity of BL in the intermediate energy range of 288.5–293.5 eV, situated between the $\pi^{*}$ and $\sigma^{*}$ resonances. This pre-$\sigma^{*}$ intensity enhancement is less dominant in the GL simulation and therefore constitutes a spectroscopic fingerprint of the modified electronic structure induced by SiC substrate hybridization. Relative to the GL (Figure~\ref{fig:figure2}b), the calculated $\pi^*$ resonance of the C$_{\mathrm{NB}}$ sites in the BL (Figure~\ref{fig:figure2}f) is considerably broader and weaker. This arises mainly from the structural disorder of the BL, which modifies both the lineshape and the chemical shift of the $\pi^*$ resonance from site to site. Although the calculation overestimates this effect, the
experimental spectra also show a lower $\pi^*$ intensity in the BL than in the GL, together with a weak additional intensity on its low-energy side (Figure~\ref{fig:figure1}d and Supporting Information, Figure~S1).
Inspection of the simulated $\mathbf{C}_{\mathrm{B}}$ spectrum further corroborates this interpretation: the spectral weight of the $\mathbf{C}_{\mathrm{B}}$ site is significantly dominated by $\sigma^{*}$ resonances, with a marked suppression of the $\pi^{*}$ feature, providing direct spectroscopic evidence for the conversion from $sp^{2}$ to $sp^{3}$ hybridization driven by covalent bond formation with the underlying Si atoms. Importantly, the $\textbf{q}_{oop}$ channel of the $\mathbf{C}_{\mathrm{NB}}$ and $\mathbf{C}_{\mathrm{B}}$ simulations also exhibits a non-negligible contribution within the 288.5--293.5~eV energy window (Figure~\ref{fig:figure2}(g)), demonstrating that the intermediate-energy intensity receives contributions from both the $\textbf{q}_{oop}$ and $\textbf{q}_{ip}$ momentum channels. This observation necessitates a rigorous comparison with the $p$DOS projected onto the relevant orbital symmetries, in order to unambiguously identify the electronic states responsible for the spectral intensity enhancement between the $\pi^{*}$ and $\sigma^{*}$ resonances and to disentangle the competing contributions of substrate-induced hybridization from intrinsic GL features.

\begin{figure}
  \centering
  \includegraphics[width=\columnwidth]{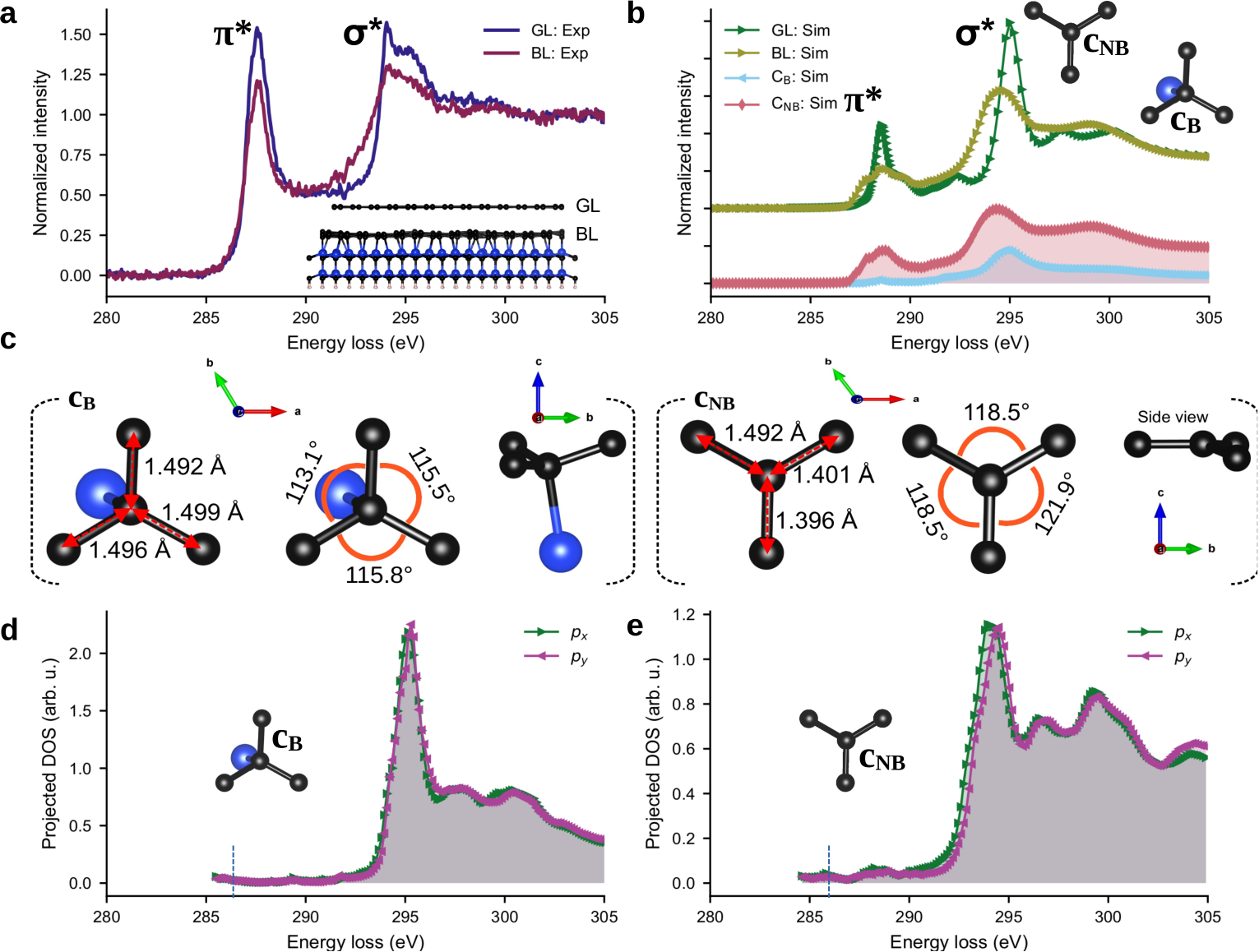}
  \caption{\textbf{Orbital Reconstruction and Broken Symmetry.} (a) Experimental C \textit{K}-edge ELNES spectra acquired from the graphene layer (GL) and buffer layer (BL). (b) Simulated C \textit{K}-edge ELNES spectra for the GL and BL. The BL spectrum is decomposed into contributions from the $\mathbf{C}_{\text{NB}}$ and $\mathbf{C}_{\text{B}}$ sites, with the total spectral response reproduced by a linear combination of 25\% $\mathbf{C}_{\text{B}}$ and 75\% $\mathbf{C}_{\text{NB}}$ character. (c) In-plane bond angles and bond lengths at the $\mathbf{C}_{\text{B}}$ and 
$\mathbf{C}_{\text{NB}}$ sites extracted from the structural model. Partial densities of states projected onto the (d) $\mathbf{C}_{\text{B}}$ and (e) 
$\mathbf{C}_{\text{NB}}$ sites reveal the lifting of orbital degeneracy arising from the broken structural symmetry at the $\mathbf{C}_{\text{NB}}$ sites. Dashed line denotes the Fermi level ($E_{\text{F}}$).}\label{fig:figure3}
\end{figure}

To identify the origin of the spectral features observed at the C $K$-edge, the simulated BL spectrum was obtained by averaging the calculated site-resolved spectra over all carbon atoms in the BL, grouped into their inequivalent $\mathbf{C}_{\mathrm{B}}$ and $\mathbf{C}_{\mathrm{NB}}$ sites, constituting approximately 25\% and 75\% of the carbon atoms (see Supporting Information, Methods), consistent with the known structural periodicity of the $(6\sqrt{3} \times 6\sqrt{3})R30^{\circ}$ reconstruction. As shown in Figures~\ref{fig:figure3}a,b, the simulated spectrum reproduces the experimental C $K$-edge, capturing both the pronounced suppression of the $\pi^{*}$ resonance and the redistribution of spectral weight into the $\sigma^{*}$ region relative to the GL reference. The simulation additionally exhibits a minor high-energy shoulder on the $\pi^{*}$ peak of the GL, arising purely from the $\textbf{q}_{oop}$ component, that is not clearly resolved in the experimental spectrum. In the BL, the $\pi^{*}$--$\sigma^{*}$ separation is reduced by approximately 0.5~eV with respect to the GL (Figure~\ref{fig:figure3}b), so that the $\sigma^{*}$ resonance is not only broadened but its maximum is also shifted to lower energy. Since the strain is largely relieved in the $(6\sqrt{3} \times 6\sqrt{3})R30^{\circ}$ model, we attribute this reduction to the corrugation of the BL and to the coexistence of $\mathbf{C}_{\mathrm{NB}}$ and $\mathbf{C}_{\mathrm{B}}$ sites, whose $\sigma^{*}$ contributions are offset from each other by slightly less than 1~eV. The overall agreement with experiment is consistent with approximately 25\% of the carbon atoms in the BL being covalently bonded to the Si-terminated SiC substrate, with the remaining 75\% retaining a non-bonded, $sp^{2}$-like coordination, in line with the 25:75 $\text{C}_{\text{B}}$:$\text{C}_{\text{NB}}$ interfacial bonding ratio obtained independently from DFT structural relaxation\cite{Radtke2025}. Nonetheless, some discrepancies between the experimental and calculated spectra remain and can be ascribed to the implementation of the core hole, which underestimates the $\pi^{*}$ intensity, as well as to electron-beam channeling effects, which induce deviations from the geometric weights of $\textbf{q}_{ip}$ and $\textbf{q}_{oop}$.

Having quantitatively established the $sp^{2}$/$sp^{3}$ site distribution within the BL, we proceeded to examine the $p$DOS for the GL, and the BL (i.e., $\mathbf{C}_{\text{B}}$ and $\mathbf{C}_{\text{NB}}$) sites individually. In the GL, the three-fold rotational symmetry of the honeycomb lattice enforces strict degeneracy of the in-plane $p_{x}$ and $p_{y}$ orbitals, as reflected in their identical $p$DOS profiles. A direct comparison of the $p_{x}$ and $p_{y}$ contributions at the $\mathbf{C}_{\text{B}}$ and $\mathbf{C}_{\text{NB}}$ sites of the BL reveals that this degeneracy is selectively and inequivalently lifted across the two populations upon covalent pinning to the SiC surface, fundamentally reorganizing the electronic structure relative to that of epitaxial GL. At $\mathbf{C}_{\text{B}}$ sites, which represent the minority (25\%) population, the 25:75 $\text{C}_{\text{B}}$:$\text{C}_{\text{NB}}$ ratio implies that each $\mathbf{C}_{\text{B}}$ atom is coordinated by an average of three $\mathbf{C}_{\text{NB}}$ nearest neighbors. The out-of-plane covalent C--Si bond to the underlying SiC substrate induces partial $sp^{3}$ rehybridization and locally suppresses aromatic delocalization, converting the three in-plane bonds into nearly equivalent single C--C bonds ($\sim$1.50~\AA), as shown in Figure~\ref{fig:figure3}c,d. Because all three neighbors are chemically equivalent, the local threefold symmetry is preserved and the $p_x$/$p_y$ degeneracy remains essentially intact. In contrast, each $\mathbf{C}_{\text{NB}}$ site is, on average, coordinated by two $\mathbf{C}_{\text{NB}}$ and one $\mathbf{C}_{\text{B}}$ nearest neighbor (with some local deviations from this idealized count). The two bonds to $\mathbf{C}_{\text{NB}}$ neighbors retain a graphene-like aromatic character (1.40--1.42~\AA), whereas the bond to the $\mathbf{C}_{\text{B}}$ neighbor, whose aromaticity is suppressed by its bonding to Si, is elongated and inequivalent ($\sim$1.49~\AA), breaking the local threefold rotational symmetry, as evident from the structural models in Figure~\ref{fig:figure3}c. This asymmetric bonding environment, together with an out-of-plane displacement markedly smaller than that at $\mathbf{C}_{\text{B}}$ sites, results in a pronounced splitting of the $p_x$- and $p_y$-derived states, as shown in Figure~\ref{fig:figure3}e. 

This lifting of orbital degeneracy at the $\mathbf{C}_{\text{NB}}$ sites reflects the anisotropic strain and broken local symmetry that arise from the proximity of covalently bonded $\mathbf{C}_{\text{B}}$ neighbors within the $(6\sqrt{3} \times 6\sqrt{3})R30^{\circ}$ superstructure, even though the $\mathbf{C}_{\text{NB}}$ atoms themselves do not form direct bonds with the substrate. This result is particularly significant, as it demonstrates that the electronic perturbation associated with substrate bonding extends beyond the immediately bonded atomic sites and propagates non-locally across the carbon network.

To further test this interpretation, and to expose the orbital symmetry lifting in a more transparent structural setting, simulations were also performed on a deliberately simplified $(\sqrt{3} \times \sqrt{3})R30^{\circ}$ model \cite{mattausch2007ab}. Although this structure departs substantially from the realistic $(6\sqrt{3} \times 6\sqrt{3})R30^{\circ}$ superstructure — most notably in its absence of significant corrugation at both the GL and BL (Figure~\ref{fig:figure2}d and see Supporting Information, Figure S5) — it nonetheless provides a valuable platform for isolating and verifying the key physical mechanisms identified in the full supercell calculations. A comparison of the simulated GL (Figure~S6), $\text{C}_{\text{B}}$ (Figure~S7), $\text{C}_{\text{NB}}$ (Figure~S7), and total BL (Figure~S8) spectra against the corresponding experimental data, obtained using the simplified $(\sqrt{3} \times \sqrt{3})R30^{\circ}$ model structure, is provided in the Supporting Information. 
As anticipated, the simplified structure reproduces the principal experimental spectral features, although the $\pi^{*}$--$\sigma^{*}$ splitting is less well resolved than in the $(6\sqrt{3} \times 6\sqrt{3})R30^{\circ}$ calculations. This is likely a consequence of the strain (around~8\%) imposed on the carbon layer, which is forced to match the SiC lattice parameter in the simplified structure, whereas in the large realistic cell it is almost completely relieved.

Crucially, however, the $p$DOS results obtained from the simplified model are in excellent agreement with those of the supercell: the $p_x$/$p_y$ degeneracy remains essentially intact at the C$_{\mathrm{B}}$ sites (Figure~3(d) and Supporting Information, Figure~S9), while a pronounced splitting of the $p_x$- and $p_y$-derived states is recovered at the C$_{\mathrm{NB}}$ sites (Figure~3(e) and Supporting Information, Figure~S10), consistent with the symmetry reduction imposed by the surrounding covalently bonded network. The convergence of results across two structurally distinct models demonstrates that the observed orbital degeneracy lifting is a robust, physically transferable feature of the BL electronic structure. This degeneracy lifting has a direct spectroscopic consequence: splitting of the $p_x$- and $p_y$-derived states at the C$_{\mathrm{NB}}$ sites shifts part of the in-plane $\sigma^*$ spectral weight to lower energy, contributing to the low-energy $\sigma^*$ shoulder observed experimentally in the BL. It is not, however, the sole origin of this shoulder: the C$_{\mathrm{B}}$ and C$_{\mathrm{NB}}$ $\sigma^*$ contributions are offset by slightly less than 1~eV (Figure~~\ref{fig:figure3}b), and the reduced $\pi^*$--$\sigma^*$ separation arising from BL disorder likewise moves the spectral weight to lower energy. The shoulder thus reflects the combined effect of these three contributions.

Furthermore, irrespective of the model, the $\sigma^*$ spectral region at both C$_{\mathrm{NB}}$ and C$_{\mathrm{B}}$ site types remains predominantly of $s$, $p_x$, and $p_y$ character.
At the C$_{\mathrm{B}}$ sites, however, the $p_z$ orbitals contribute significantly to the $\sigma^*$ region (Supporting Information, Figures~S9 and S11(b)), as expected from the formation of the C--Si bond. The $p_z$ contribution is smaller than the $p_x$ and $p_y$ contributions at the $\sigma^*$ maximum (294.5~eV), but it dominates between 290 and 293~eV, at the energy of the low-energy shoulder of the $\sigma^*$ resonance. This shift of the $p_z$ spectral weight towards low-energy is a fingerprint of the covalent interaction with the underlying Si atoms. Its distribution differs markedly from that of a perfect $sp^3$ hybridization, for which the $p_x$, $p_y$ and $p_z$ contributions would be equivalent.
This is consistent with a partial, rather than complete, $sp^3$ rehybridization at this site. By contrast, the $p_z$ contribution is much weaker at the C$_{\mathrm{NB}}$ sites (Figure~S11(c)), as expected for carbon atoms that are not bonded to the substrate and therefore retain a predominantly $sp^2$-like configuration.
The site-resolved $p$DOS analysis thus reveals that the electronic structure reorganization within the BL cannot be attributed solely to local $sp^{3}$ rehybridization at the covalently bonded $\mathbf{C}_{\mathrm{B}}$ sites; rather, it emerges from a collective symmetry breaking that propagates across the extended carbon network via non-local modulation of in-plane orbital degeneracy --- a fundamentally cooperative electronic response to substrate interaction.

The inclusion of $p_{z}$ orbital contributions reveals a further consequential departure from graphenic behavior that directly extends this symmetry-breaking picture. At $\mathbf{C}_{\text{B}}$ sites, the formation of covalent bonds with the underlying Si atom drives significant $sp^{3}$-like hybridization, as evidenced by the suppression and concomitant downward energy shift of the intensity projected $p_{z}$ in the $p$DOS (Supporting Information, Figure~S11), localizing the $p_{z}$ electron in a low-energy bonding state and withdrawing it from the itinerant $\pi$ conduction network. At $\mathbf{C}_{\mathrm{NB}}$ sites, however, the robust $p_z$ intensity at higher energies (Supporting Information, Figures~S9,~S10,~and~S11) confirms the $\pi$ bonding, despite a symmetry-broken local environment induced by inequivalent C–C bond lengths. The consequences are twofold: the conducting $\pi$ channel is depleted at bonded sites, while the induced symmetry breaking propagates laterally to neighboring $\mathbf{C}_{\mathrm{NB}}$ sites, lifting the degeneracy of their frontier orbitals --- consistent with the $p_{x}$/$p_{y}$ splitting identified in the $p$DOS analysis above. The occurrence of these two mechanisms --- $\pi$-electron localization via $sp^{3}$ rehybridization at $\mathbf{C}_{\text{B}}$ sites and site-specific orbital degeneracy lifting at $\mathbf{C}_{\text{NB}}$ sites --- constitutes a peculiar electronic structure characteristic of the graphene BL--SiC interface. Both mechanisms lift the degeneracy of the graphene sublattices, a necessary condition for opening a bandgap at the Dirac point.\cite{zhou2007,Zhao2024} This sublattice symmetry breaking, driven by the selective covalent bonding of $\mathbf{C}_{\text{B}}$ sites to the underlying SiC substrate, has been experimentally evidenced by the absence of dispersive $\pi$-bands in angle-resolved photoemission spectroscopy measurements of the BL and the emergence of localized non-dispersive states below the Fermi energy,\cite{zhou2007,nair2017} and distinguishes the electronic character of the BL fundamentally from the quasi-free-standing graphene character of the epitaxial overlayer.

\section{Conclusion}

In summary, monochromated core-loss EELS combined with \textit{ab initio} calculations have been employed to resolve the electronic structure of the epitaxial graphene BL on SiC at atomic plane resolution. The C $K$-edge fine structure of the BL is markedly distinct from that of quasi-free-standing graphene, with the weakening of the $\pi^{*}$ resonance and redistribution of spectral weight into the $\sigma^{*}$ region quantitatively reproduced by a 25\% bonded and 75\% non-bonded \textbf{C} site decomposition consistent with the surface reconstruction. Beyond structural confirmation, the site-resolved partial density of states analysis provides direct electronic insight into the BL's orbital symmetry. Covalent bonding at $\mathbf{C}_{\text{B}}$ sites drives $sp^{3}$-like rehybridization that localizes $p_{z}$ electrons into low-energy bonding states, depleting the $\pi$ conduction network, while the induced symmetry breaking propagates to $\mathbf{C}_{\text{NB}}$ sites through a selective lifting of $p_{x}$/$p_{y}$ orbital degeneracy. These findings, accessible only to high-spatial-resolution core-level spectroscopic techniques, underscore the unique capability of sub-Å core-loss EELS to disentangle orbital-level electronic reconstruction at buried graphene interfaces, providing atomic-scale insight into the origin of the buffer-layer bandgap and suggesting that substrate hybridization represents a viable route to tailoring the electronic structure of epitaxial graphene.

\begin{acknowledgement}
VKP and MB acknowledge funding from Agence Nationale de la Recherche (Grant ANR-22-CE29-0019). The EELS experiments were performed at SuperSTEM, the EPSRC national facility for advanced electron microscopy (through grants EP/V036432/1 and EP/W021080/1). The authors thank Orsay Physics for lamella preparation by focused ion beam milling. M.B. acknowledges the School of Chemical and Process Engineering at the University of Leeds for a Visiting Associate Professorship.
\end{acknowledgement}

%
% ...

\bibliography{references}

\begin{tocentry}
  \includegraphics[width=83mm,height=47mm,keepaspectratio]{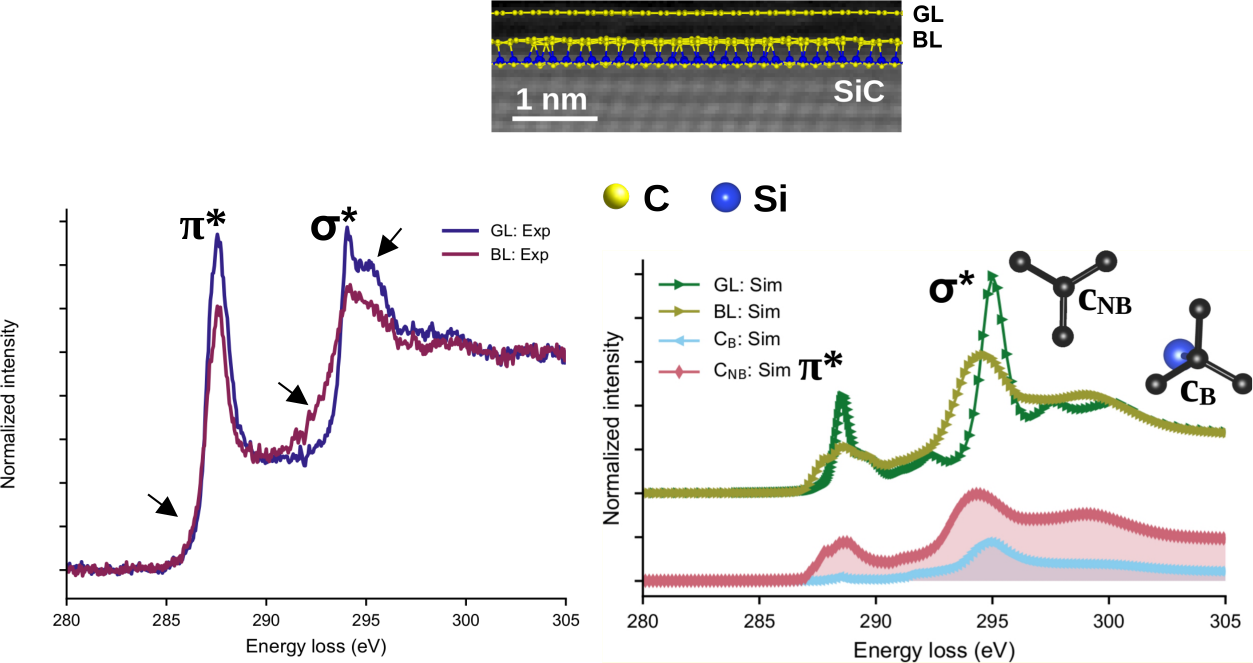}
\end{tocentry}


\clearpage

\clearpage

\clearpage

\clearpage
\includepdf[pages=-]{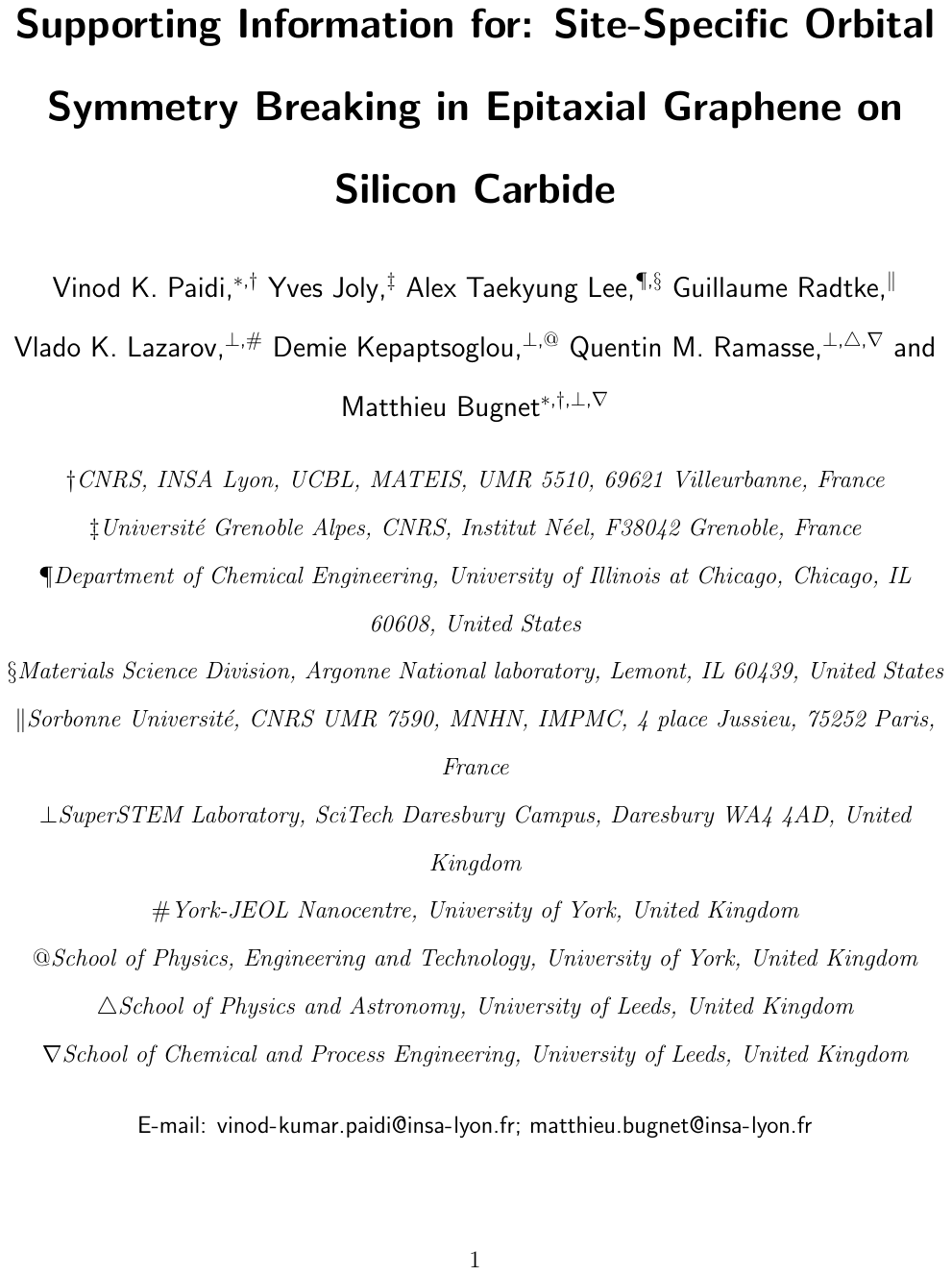}
\end{document}